\documentclass[
    ,final            
  ]
  {aipproc}
\usepackage[usenames,dvipsnames]{color} 
\usepackage{multirow}
\usepackage{amsmath}
\usepackage{mciteplus}

\layoutstyle{6x9}

\newcommand{\ii}{i}                    
\newcommand{\de}{d}                    

\newcommand{\bm}{\boldmath}

\begin{document}

\title{What can we learn from TMD measurements?\footnote{
Authored by Jefferson Science Associates, LLC under U.S. DOE Contract
No. DE-AC05-06OR23177.  
The U.S. Government retains a non-exclusive, paid-up, irrevocable, world-wide
license to publish or  
reproduce this manuscript for U.S. Government purposes.}}


\author{Alessandro Bacchetta}{
  address={Theory Center, Jefferson Lab, 12000 Jefferson Ave, Newport News, VA
    23606, USA} 
}

\begin{abstract}
Transverse-momentum-dependent parton distribution and fragmentation functions
describe the partonic structure of the nucleon in a
three-dimensional momentum space. They are subjects of 
flourishing theoretical and experimental activity. 
They provide novel and intriguing information on hadronic
structure, including evidence of the presence of partonic 
orbital angular momentum.
\end{abstract}

\maketitle


TMDs is an acronym for Transverse Momentum Distributions or Transverse
Momentum Dependent parton distribution functions, also called
unintegrated parton distribution functions. 
Most of our knowledge of the inner structure of nucleons is encoded in parton
distribution functions (PDFs). We introduce them to describe hard scattering
processes involving nucleons. 
The presence of a hard probe in these processes --- e.g., in DIS the
photon with virtuality $Q^2$ --- identifies a longitudinal direction, 
and a plane perpendicular to that, the transverse plane.
Intuitively, standard collinear PDFs describe the probability to find in a
fast-moving 
nucleon a parton with a specific
fraction of the nucleon's longitudinal momentum. 
TMDs describe also the probability that the parton has a specific transverse
momentum.
They are therefore a natural extension of standard PDFs 
from one to three dimensions in momentum
space.

Although useful from the intuition point of view, 
the probabilistic interpretation of PDFs and TMDs has some technical problems
and is not strictly needed~\cite{Collins:2003fm}. 
What is essential is that PDFs and TMDs can be
defined in a formally clear way, through the application of factorization
theorems. They reveal crucial aspects of
the  
dynamics of confined partons, they can be extracted from experimental data,
and allow us to make prediction for
hard-scattering experiments involving nucleons. In this sense, 
the information contained in TMDs is as important as that contained in standard
PDFs. 

The main difference between collinear PDFs and TMDs is that the latter
do not appear in totally inclusive processes. For instance, they do not appear
in totally inclusive DIS, but they are needed when semi-inclusive DIS is
studied and the transverse momentum of one outgoing hadron, $P_{h \perp}$, 
is measured. They
are necessary to describe Drell--Yan processes when the transverse
momentum of the virtual photon, $q_T$, is measured.

Factorization for processes involving TMDs has been worked out explicitly at
leading twist (twist 2) and one-loop order and argued to hold at all
orders~\cite{Collins:1981uk,Ji:2004wu}.  
For instance, in unpolarized semi-inclusive DIS 
we can measure the structure function $F_{UU,T}$, which in
the region $P_{h \perp}^2 \ll Q^2$ can be expressed as~\cite{Bacchetta:2008xw}
\begin{align}
F_{UU,T} &=
\bigl| H\bigl(x \zeta^{1/2}, z^{-1}\zeta_{\smash{h}}^{1/2} , \mu_F\bigr)
\bigr|^2 \,
\sum_a x\, e_a^2
\int \de^2 \bm{p}_T\, \de^2 \bm{k}_T^{}\, \de^2 \bm{l}_T^{}\,
\nonumber \\
& \times
  \delta^{(2)}\bigl(\bm{p}_T - \bm{k}_T^{} +\bm{l}_T^{} - \bm{P}_{h \perp}^{}/z
  \bigr)\,
f_1^a(x,p_T^2;\zeta, \mu_F)\, D_1^a(z,k_T^2;\zeta_h,\mu_F)\, U(l_T^2;\mu_F) \,.
\label{FUUTconv}
\end{align}
Apart from the transverse-momentum-dependent PDFs and fragmentation functions,
the formula contains the soft factor $U$, a nonperturbative and
process-independent object.

For the specific case of unpolarized observables integrated over the azimuthal
angle of the measured transverse momentum, the analysis is usually
performed in 
$b$-space in the Collins--Soper--Sterman
framework~\cite{Collins:1984kg}.  
The
region of $P_{h \perp}^2 \gg M^2$, or $b^2\ll 1/M^2$, 
can be calculated perturbatively, but when $P_{h \perp}^2 \approx
M^2$ a nonperturbative component has to be introduced and its parameters must
be fitted to experimental data. This component is usually assumed to be a
flavor-independent Gaussian~\cite{Landry:2002ix}.

At present, especially for azimuthally-dependent structure functions,
 phenomenological analyses are often carried out using the
tree-level approximated expression 
\begin{align}
F_{UU,T} &=
\sum_a x\, e_a^2
\int \de^2 \bm{p}_T\, \de^2 \bm{k}_T^{}\,
  \delta^{(2)}\bigl(\bm{p}_T - \bm{k}_T^{} - \bm{P}_{h \perp}^{}/z
  \bigr)\,
f_1^a(x,p_T^2)\, D_1^a(z,k_T^2)\,.
\label{FUUTconv_tree}
\end{align}
Also in this case, the transverse-momentum dependence of the
partonic functions is assumed to be a flavor-independent Gaussian~\cite{D'Alesio:2004up}. The
tree-level approximation and the Gaussian
assumption are known to be inadequate at $P_{h \perp}^2
\gg M^2$, but they could still effectively describe the physics at $P_{h \perp}^2
\approx M^2$. Especially for low-energy experiments, this is where the bulk of
the data is.

The definition of quark TMDs is~\cite{Collins:2003fm,Ji:2004wu} 
(taking the example of the fully
unpolarized distribution of a quark with flavor $a$)
\begin{equation}  
f_1^a
(x,p_T^2; \zeta, \mu_F)= \int 
        \frac{\de \xi^- \de^2 \bm{\xi}_T}{(2\pi)^{3}}\; 
 e^{\ii p \cdot \xi}\,
       \langle P|\bar{\psi}^a(0)\,
{\cal L}^{v \dagger}_{(\pm\infty,0)}\,
\gamma^+
{\cal L}^{v}_{(\pm\infty,\xi)}\,
\psi^a(\xi)|P \rangle \bigg|_{\xi^+=0}.
\label{e:phi} 
 \end{equation} 
The Wilson lines, ${\cal L}$, guarantee the gauge invariance of the TMDs. They
depend on the gauge vector $v$ and 
contain also components at infinity running in the transverse direction.
A remarkable property of TMDs is that the detailed shape of the Wilson line is
process-dependent. This immediately leads to the conclusion that TMDs are not
universal. However, the situation is not hopeless and the predictive power of
TMD factorization is not completely destroyed, for the following reasons
\begin{itemize}
\item{For transverse-momentum-dependent fragmentation functions, the shape of
    the Wilson line appears to have no influence on physical
    observables~\cite{Metz:2002iz,*Collins:2004nx,*Yuan:2008yv,*Gamberg:2008yt,*Meissner:2008yf}.}  
\item{In SIDIS and Drell--Yan, the difference between the Wilson line consists
  in a simple direction reversal and leads to calculable effects, namely a
  simple sign reversal of all T-odd TMDs~\cite{Collins:2002kn}.}
\item{In hadron-hadron collisions to hadrons, standard universality cannot be
    applied. It is however conceivable that only 
     a manageable number of TMDs with distinct Wilson lines are needed, 
     preserving part of the
     predictive power of the formalism~\cite{Collins:2007nk,*Vogelsang:2007jk}.}
\item{If we consider specific transverse-momentum-weighted observables instead
    of unintegrated observables, it should be possible to obtain factorized
     expressions in terms of transverse moments of TMDs multiplied by
     calculable, process-dependent
     factors~\cite{Bacchetta:2005rm,*Bomhof:2006ra,*Bacchetta:2007sz}.} 
\end{itemize}

Our understanding of TMDs and their extraction from data has made giant steps
in the last years, thanks to new theoretical ideas and experimental
measurements. In the near future, more experimental data are expected from
HERMES, COMPASS, BELLE and JLab.

When the spin of the nucleon and that of the quark are taken into account,
eight twist-2 functions can be introduced. They are listed in 
Tab.~\ref{t:TMDs-tw2}. 
As with collinear PDFs, extracting TMDs calls for global fits to
semi-inclusive DIS, Drell--Yan, and $e^+e^-$-annihilation data. Care has to be
taken when considering the peculiar universality properties of TMDs. 
At the moment, we have some information only about the
two functions in the first column of the table: $f_1$ (unpolarized function) and
$f_{1T}^{\perp}$ (Sivers function).

\begin{ltxtable}
\begin{center}
  \includegraphics[height=3.5cm]{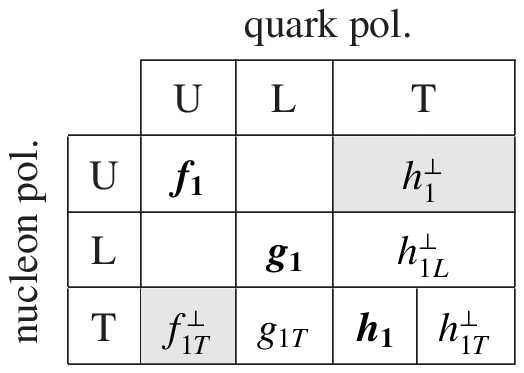}
\end{center}
\caption{Twist-2 transverse-momentum-dependent distribution functions.
The U,L,T correspond to unpolarized, longitudinally polarized and transversely
polarized nucleons (rows) and quarks (columns). Functions in boldface survive
transverse momentum integration. Functions in gray cells are T-odd.} 
\label{t:TMDs-tw2}
\end{ltxtable}

\begin{figure}
\begin{tabular}{cc}
  \includegraphics*[height=5cm]{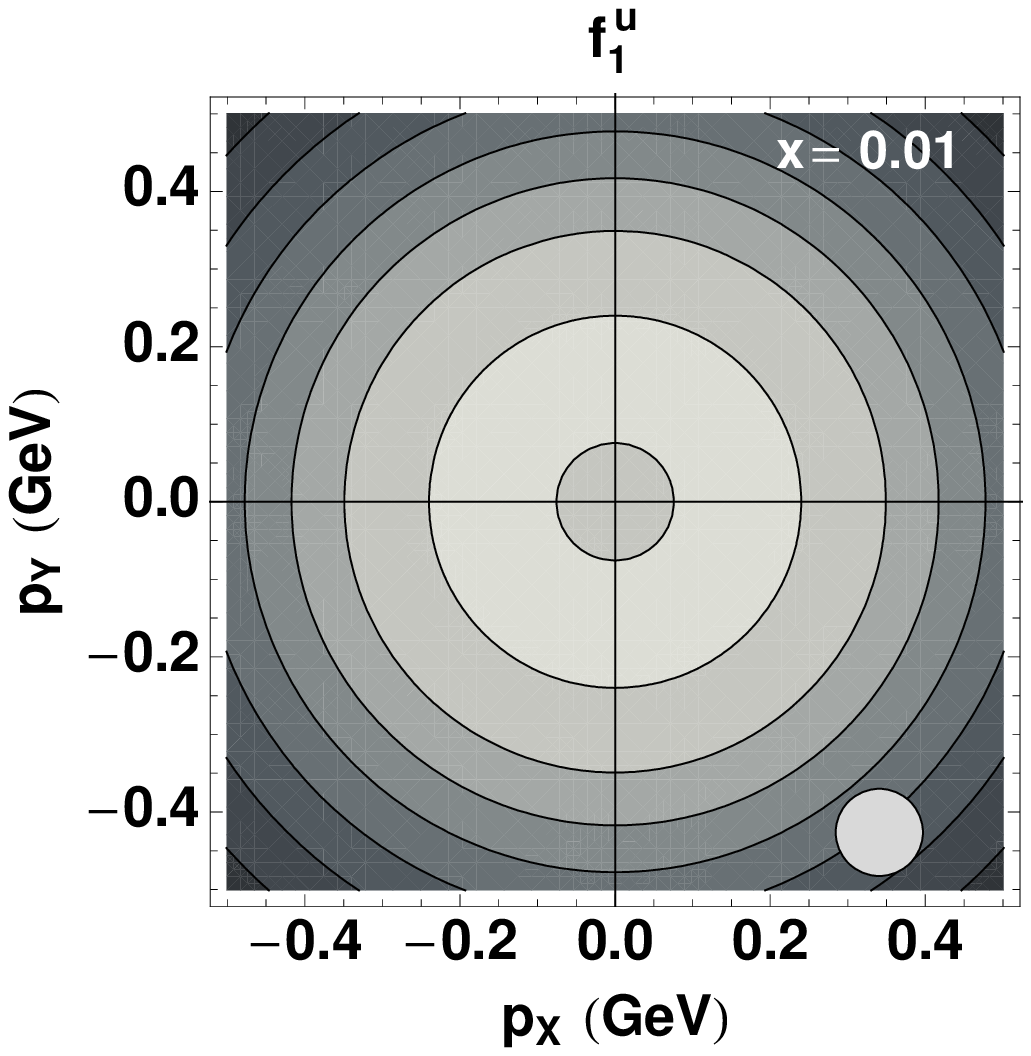}
&
  \includegraphics[height=5cm]{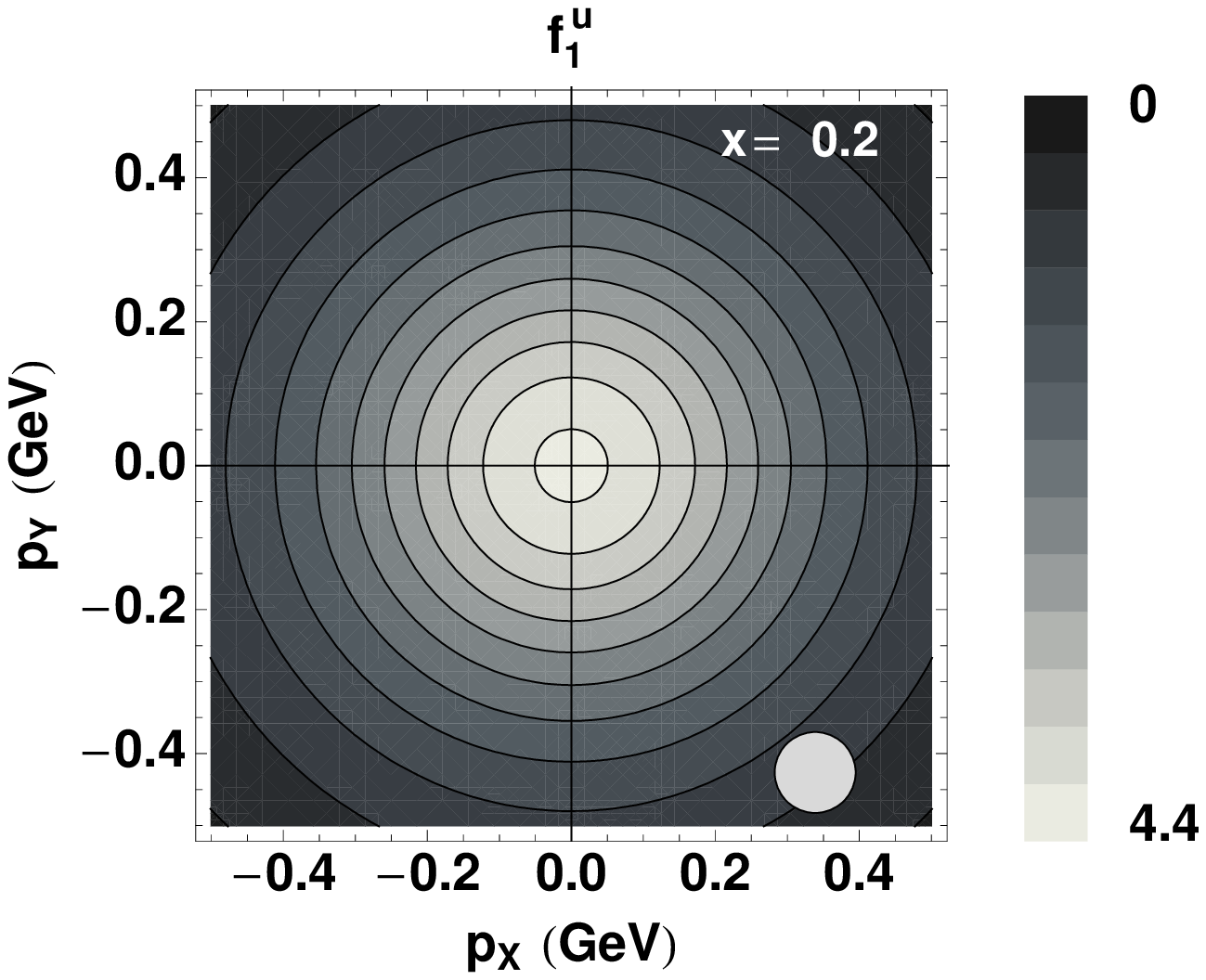}
\\
  \includegraphics*[height=5cm]{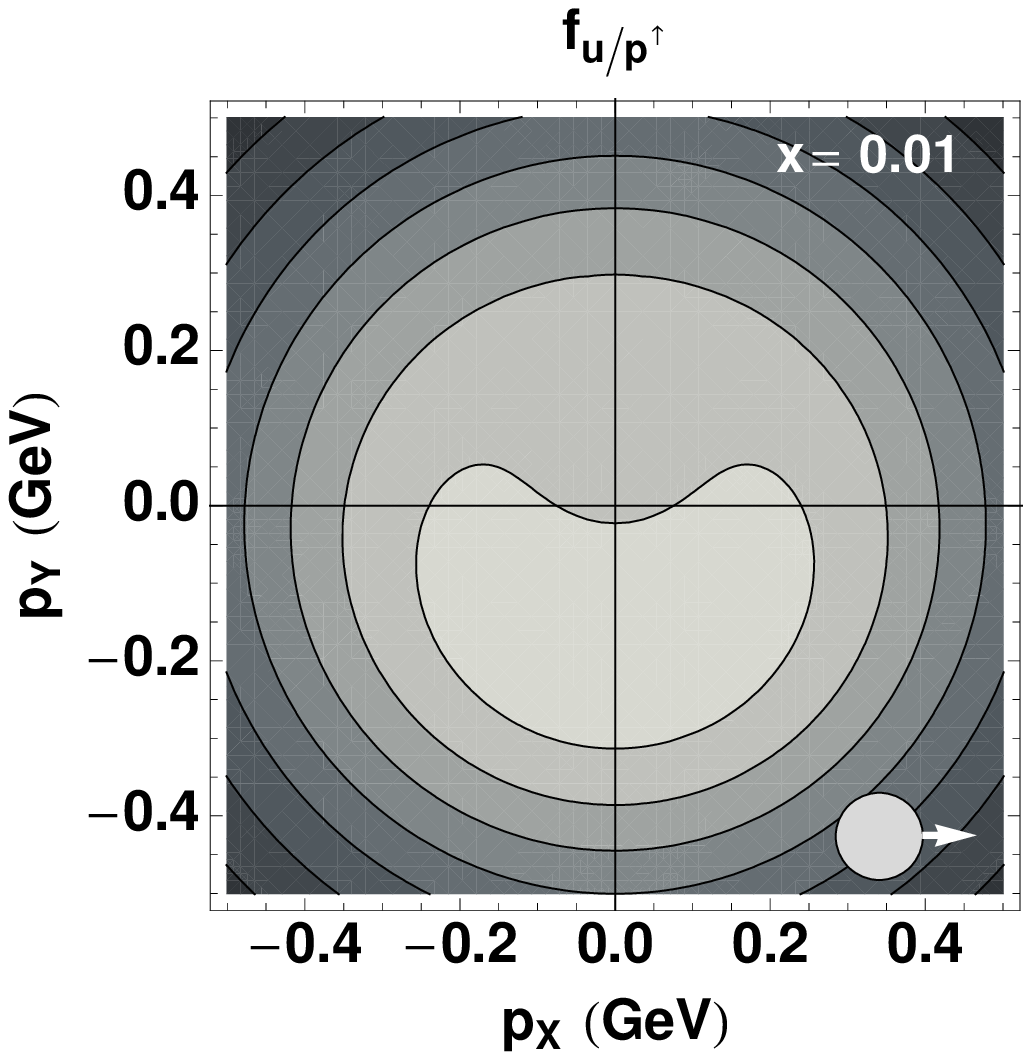}
&
  \includegraphics[height=5cm]{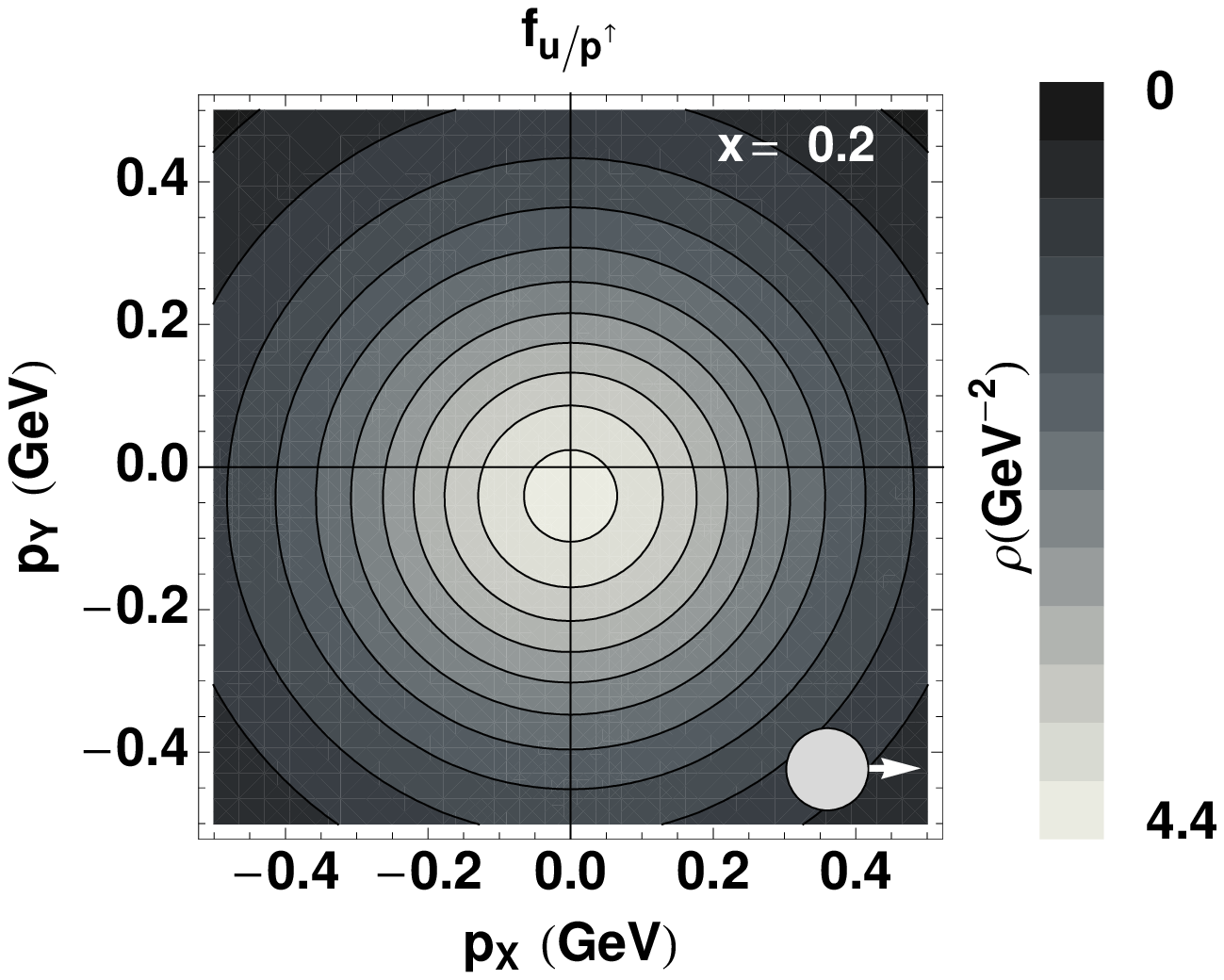}
\end{tabular}
  \caption{Momentum-space tomographic ``images'' of the up quarks in a nucleon
    obtained from a model calculation of TMDs~\cite{Bacchetta:2008af}. 
    The circle with the arrow
    indicates the nucleon and its spin orientation. The distortion in the
    lower panels is due to the Sivers function. In the future, it should be
    possible to reconstruct these images from experimental data.}
\label{f:tomo}
\end{figure}

Ultimately, the knowledge of TMDs should allow us to build tomographic images
of the inner structure of the nucleon in momentum space, complementary to the
impact-parameter space tomography that can be achieved by studying generalized
parton distribution functions (GPDs). An example of tomographical images of
the nucleon based on a model calculation of TMDs~\cite{Bacchetta:2008af} is
presented in 
Fig.~\ref{f:tomo}.  

\renewcommand{\arraystretch}{1}

TMDs measurements should allow us to address 
some intriguing questions, e.g., 
\begin{itemize}
\item{Are there differences between the TMDs of different quark flavors (and
    of gluons)? We know that collinear PDFs are different, not only in
    normalization, but also in shape. We can expect that also the transverse
    momentum distribution is different. See Ref.~\cite{Mkrtchyan:2007sr} for
    an example of an experimental analysis of this issue.}
\item{How does the transverse momentum dependence change with $x$? 
Such a  dependence has already been introduced to describe data at low
    $x$~\cite{Landry:2002ix}.} 
\item{Does the transverse momentum dependence of fragmentation functions
    change for different quark flavors and different produced hadrons?}
\item{Are there reasons to abandon a Gaussian ansatz? We know that
  this assumption fails at high transverse momentum, but there are no
  compelling reasons to take a Gaussian shape even for the
  low-transverse-momentum, nonperturbative region.}
\end{itemize}
 
\begin{figure}
  \includegraphics[height=4.5cm]{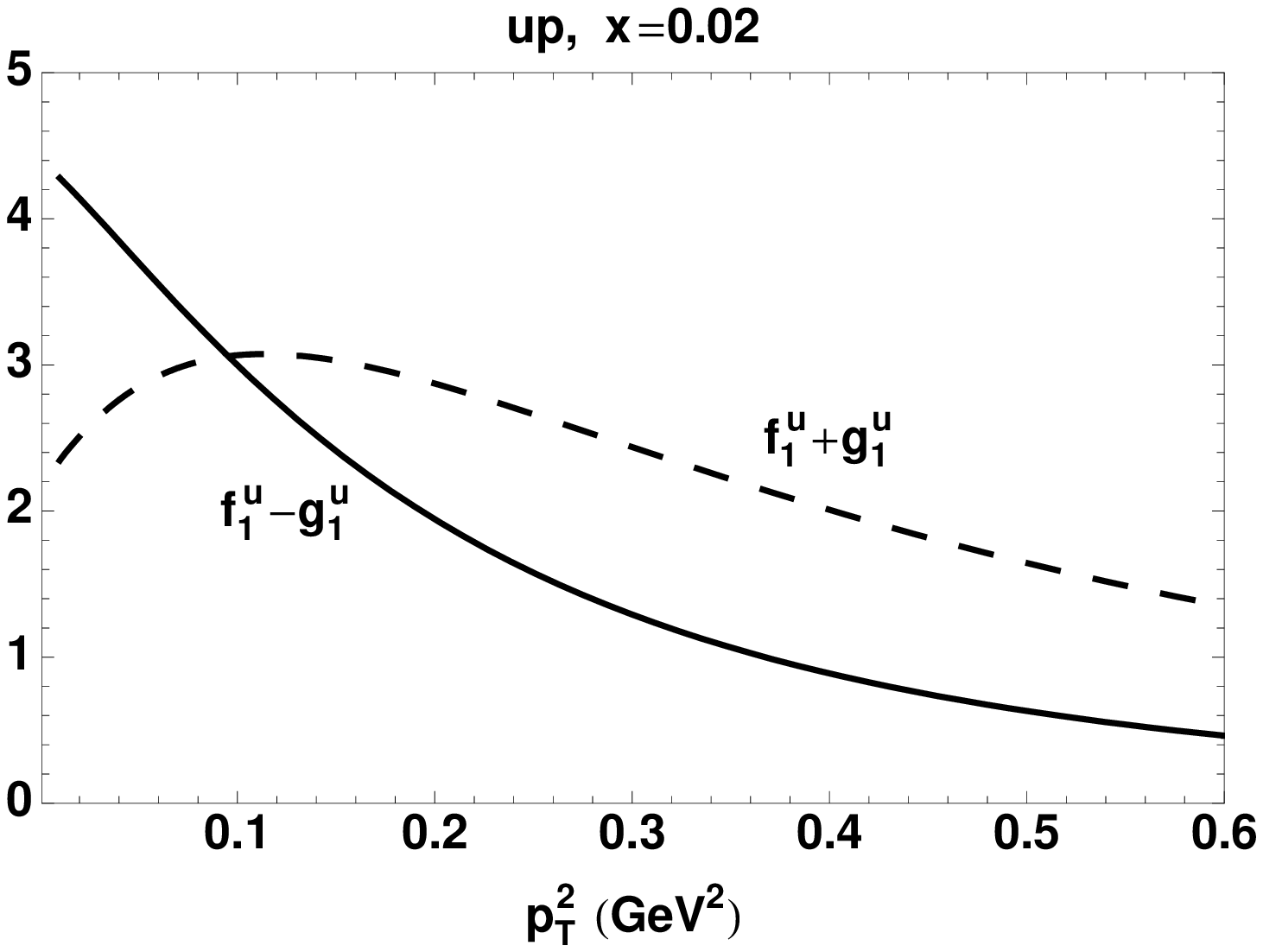}
\hfil
  \includegraphics[height=4.5cm]{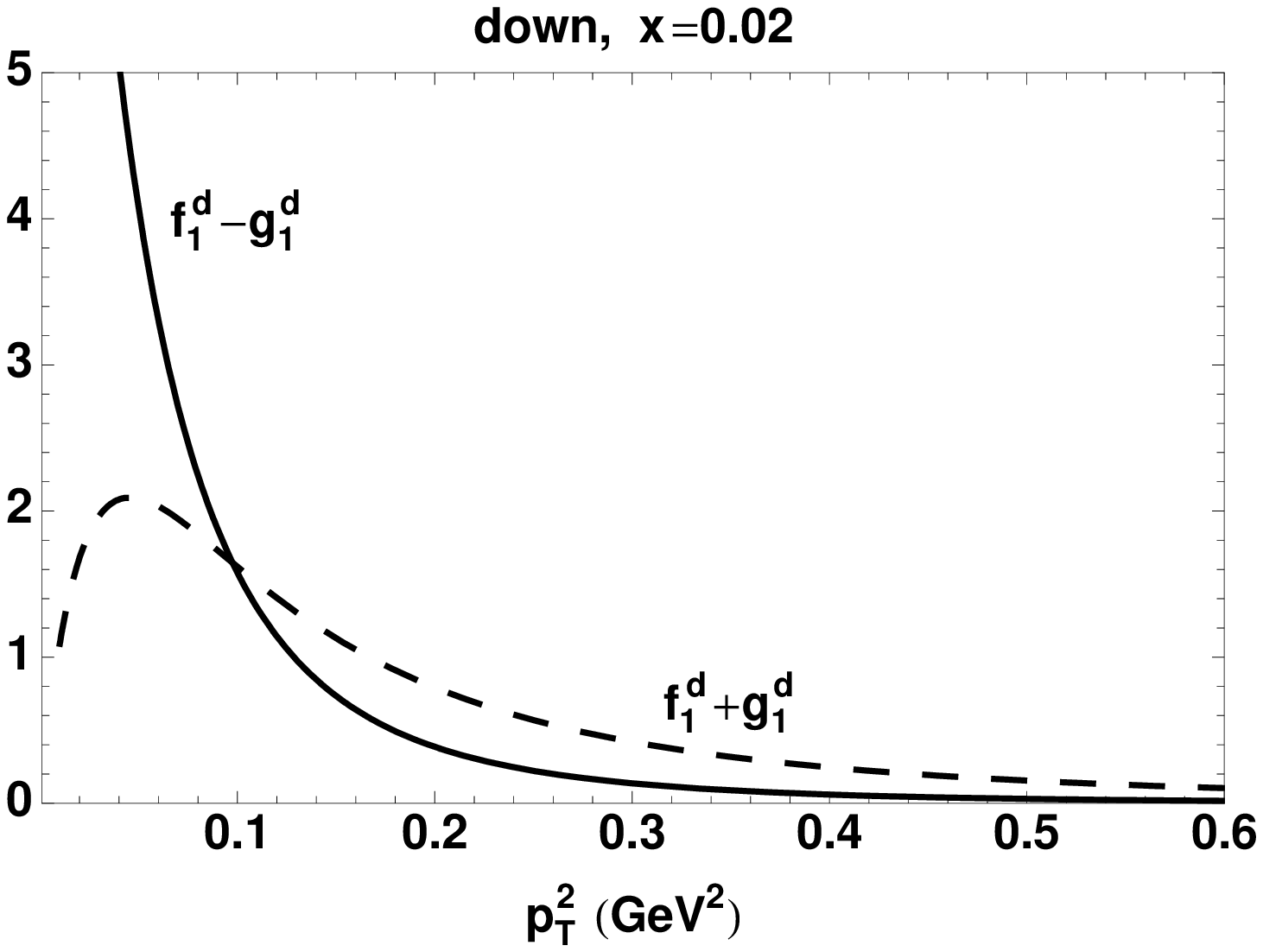}
  \caption{An illustration of how the presence of orbital angular momentum can
  influence the shape of TMDs. The model calculation shows different
  combinations of the $f_1$ and $g_1$ TMDs for $u$ and $d$ quark at
  $x=0.02$. The downturns for $p_T^2 \to 0$ are due to the presence of orbital
  angular momentum.}
\label{f:f1g1}
\end{figure}

The last item of the list is connected also to another fundamental issue that
makes TMDs interesting, i.e., the observation of partonic orbital angular
momentum. In nonrelativistic quantum mechanics, it is well known that
wavefunctions with orbital angular momentum vanish at zero momentum. This is a
general statement independent of the specific potential in which the
wavefunction is computed. This feature is reflected also in TMDs:
contributions from partons with nonzero angular momentum have to vanish at
zero transverse momentum (and therefore 
cannot be described by a simple Gaussian). In
general, a downturn of a TMD going to zero transverse momentum can signal the
presence of nonzero orbital angular momentum. While this effects could barely
be visible in unpolarized TMDs, certain combinations of polarized TMDs could
filter out more clearly the configurations with nonzero orbital angular
momentum. Fig.~\eqref{f:f1g1} shows an example of this phenomenon, 
using a model calculation for illustration purposes.

Apart from the details of their shape, all the TMDs that are not boldface
in Tab.~\ref{t:TMDs-tw2} vanish in the absence of
orbital angular momentum due to angular momentum conservation. Measuring any
one of them to be nonzero is already an unmistakable indication of the
presence of partonic orbital angular momentum. We know already from other
sources (in particular the measurement of nucleons' anomalous magnetic
moments) that partonic orbital angular momentum is not zero, however TMDs have
the advantage that they can be flavor-separated and that they are $x$
dependent.  Thus, they allow us to say if
orbital angular momentum is present for each quark flavor and for gluons, and 
at each
value of $x$.

If stating that a fraction of partons have nonzero orbital
angular momentum is relatively simple, 
it is not easy to make a quantitative estimate
of the net partonic orbital angular momentum using TMDs. 
Any statement in this direction is bound to be model-dependent.
Generally speaking, TMDs have to be computed in a model and the parameters of
the 
model have to be fixed to reproduce the TMDs extracted from data.  
Then, the
total orbital angular momentum can be computed in the model. 
Unfortunately, it is 
possible that two models describe the data equally well, but
give two different values for the total orbital angular momentum.

As an example of a procedure of this kind, 
let us take the measurement of the Sivers
function. We know that the proper way to measure the quark total
angular momentum is by measuring the combination~\cite{Ji:1997ek}
\begin{equation}
J^{a}= \int_0^1 dx \, x\, 
\Bigl(H^{a}(x,0,0)+E^{a}(x,0,0) \Bigr),
\end{equation} 
where the definition of the generalized parton distribution $E$ in terms of
light-cone wavefunctions is
\begin{equation}
\begin{split} 
E(x,0,0) = \lim_{q_T \to 0} \biggl(-\frac{1}{q_x -i q_y}
\int \frac{\de^2 p_T}{16 \pi^2}\,\Bigl[&\psi^{+\ast}_+ \bigl(x, p_T\bigr)
\psi^{-}_+ \bigl(x, p_T+(1-x)q_T\bigr)
\\
& +\psi^{+\ast}_- \bigl(x, p_T\bigr)
\psi^{-}_- \bigl(x, p_T+(1-x)q_T\bigr)
\Bigr] \biggr).
\end{split}  
\end{equation} 
On the other hand, the definition of the Sivers function in terms of
light-cone wavefunctions can be written as
\begin{equation}
f_{1T}^{\perp}(x,p_T) =\frac{1}{16 \pi^3} {\rm Im}\Bigl[ \psi^{+\ast}_+ \bigl(x, p_T\bigr)
\psi^{-}_+ \bigl(x, p_T\bigr)
 +\psi^{+\ast}_- \bigl(x, p_T\bigr)
\psi^{-}_- \bigl(x, p_T\bigr)
\Bigr] .
\end{equation} 
In spite of the similarities between the two expressions and the fact that the
same light-cone wavefunctions are involved, in general there is no
straightforward connection between the Sivers function and the GPD
$E$~\cite{Meissner:2007rx}. Nevertheless, in a certain class of spectator models
it turns out that~\cite{Burkardt:2003je,*Lu:2006kt}
\begin{equation}
 f_{1T}^{\perp a}(x) = -L(x)\, E^a(x,0,0).
\end{equation} 
Exploiting this very simple relation and using for illustration purposes the
results of the Sivers function fit from Ref.~\cite{Arnold:2008ap} we obtain
\begin{equation}
\frac{E^a(x,0,0)}{E^u(x,0,0)} =\frac{f_{1T}^{\perp a}(x)}{f_{1T}^{\perp u}(x)}
= \frac{A_a}{A_u}\,\frac{f_1^a(x)}{f_1^u(x)}, 
\end{equation} 
where (error estimates do not take into account parameter correlations)
\begin{align}
\frac{A_d}{A_u} &= -1.8 \pm 0.2 ,
&
\frac{A_{\bar{u}}}{A_u} &= -1.1 \pm 0.1,
&
\frac{A_{\bar{d}}}{A_u} &=  1.3 \pm 0.2,
&
\frac{A_s}{A_u} = - \frac{A_{\bar{s}}}{A_u} &= -4.8.
\end{align} 
Although assumption-based, the above analysis shows that the measurement of
the Sivers function can
be used to give interesting constraints on the GPD $E$ and ultimately on the
amount of total orbital angular momentum for each flavor.

In summary, TMDs open new dimensions in the exploration of the partonic
structure of the nucleon. They require challenging extensions of the standard
formalism used for collinear parton distribution functions, leading 
 us to a deeper understanding of QCD. Among other things, 
they give evidence of the presence of
partonic orbital angular momentum and, with model assumptions, they can help
constraining its size. 




\bibliographystyle{aipprocM}   

\bibliography{mybiblio}

\IfFileExists{\jobname.bbl}{}
 {\typeout{}
  \typeout{******************************************}
  \typeout{** Please run "bibtex \jobname" to optain}
  \typeout{** the bibliography and then re-run LaTeX}
  \typeout{** twice to fix the references!}
  \typeout{******************************************}
  \typeout{}
 }

\end{document}